\documentclass[aps,prl,twocolumn,groupedaddress]{revtex4-1}
\usepackage{bm}
\usepackage{graphicx}
\usepackage[usenames]{color}
\usepackage{amsmath,amsfonts,amssymb}
\usepackage{hyperref}
\usepackage{ulem}
\usepackage{xcolor}

\newcommand{\ket}[1]{|{#1}\rangle}
\newcommand{\bra}[1]{\langle{#1}|}
\newcommand{\pa}{\partial}

\allowdisplaybreaks[4]

\begin{document}
\title{Bulk-edge Correspondence in the Adiabatic Heuristic Principle}

\author{Koji Kudo$^{1,2}$}
\altaffiliation[Present address: ]{Department of Physics, 104 Davey Lab, 
The Pennsylvania State University, University Park, Pennsylvania 16802, USA}
\author{Yoshihito Kuno$^1$}
\altaffiliation[Present address: ]{Graduate School of Engineering Science, 
Akita University, Akita, 010-8502, Japan}
\author{Yasuhiro Hatsugai$^{1,2}$}
\affiliation{
$^1$Department of Physics,
University of Tsukuba, Tsukuba, Ibaraki 305-8571, Japan\\
$^2$Graduate School of Pure and Applied Sciences, University of Tsukuba, 
Tsukuba, Ibaraki 305-8571, Japan
}

\date{\today}

\begin{abstract}
 Using the Laughlin's argument on a torus with two pin-holes, we numerically 
 demonstrate 
 that the discontinuities of the center-of-mass work well as an invariant of 
 the pumping phenomena during the process of the flux-attachment, trading the 
 magnetic flux for the statistical one. This is consistent with the bulk-edge 
 correspondence of the fractional quantum Hall effect of anyons. We also 
 confirm that the general feature of the edge states remains unchanged during 
 the process while the topological degeneracy is discretely changed. 
 This supports the stability of the quantum Hall edge 
 states in the adiabatic heuristic principle.
\end{abstract}

\maketitle

\paragraph{Introduction}---
Characterization of quantum matter with topological invariants is a modern 
notion in condensed matter 
physics~\cite{Thouless_PRL82,Kohmoto_Ann85,Hasan_PMP10,Qi_PMP11}. The adiabatic
deformation of gapped systems is a conceptual basis in the theory of 
topological phases beyond the 
Landau's symmetry breaking paradigm~\cite{Wen_PRB89,Wen_PMP17}. 
Meanwhile, augmented by the symmetry, this notion leads to more 
unified picture exemplified by the ``periodic table'' for topologically 
nontrivial states~\cite{Kitaev_AIP09,Schnyder_PRB08,Qi_PRB08,Ryu_IOP10} and
demonstrates the existence of rich topological phases. The adiabatic 
deformation also gives a useful way to characterize concrete models by reducing
them to simple
systems~\cite{Greiter_NPB90,Greiter_NPB92,Greiter_arXiv21,Hatsugai_JPSJ05,Hatsugai_JPSJ06,Hatsugai_IOP07,Kariyado_PRL18}.

The adiabatic heuristic argument of the quantum Hall (QH) 
effect~\cite{Greiter_NPB90,Greiter_NPB92,Greiter_arXiv21} is the historical 
example in which the adiabatic deformation has been successfully used.
The fractional QH (FQH) effect~\cite{Tsui_PRL82,Laughlin_PRL83} is a 
topological ordered phase~\cite{Wen_AP95} with fractionalized 
excitations~\cite{Haldane_PRL83,Halperin_PRL84,Arovas_PRL84}. 
Even though it is intrinsically a many-body problem of correlated electrons
unlike the integer QH (IQH) 
effect~\cite{Klitzing_PRL80,Laughlin_PRL81,Thouless_PRL82,Halperin_82}, 
the composite fermion theory~\cite{Jain_CFT_PRL89,Jain_book07} gives a unified 
scheme to describe their underlying physics: the FQH state at the filling 
factor $\nu=p/(2mp\pm1)$ with $p,m$ integers can be interpreted as the $\nu=p$ 
IQH state of the composite fermions. By continuously trading 
the external flux for the statistical one~\cite{Wilczek1_PRL82,Wilczek2_PRL82},
both states are adiabatically connected through intermediate systems of anyons 
(adiabatic heuristic 
principle~\cite{Greiter_NPB90,Greiter_NPB92,Greiter_arXiv21}). Even though the 
ground state degeneracy~\cite{Haldane_PRL85,Wen_Deg_PRB90} is wildly 
changed in the periodic geometry~\cite{Greiter_NPB92,Kudo_PRB20}, the energy 
gap remains open and its many-body Chern 
number~\cite{Niu_PRB85} 
works well as an adiabatic invariant~\cite{Kudo_PRB20}.

Generally bulk topological invariants such as the Chern number are intimately 
related to the presence of gapless edge excitations. This is the so-called 
bulk-edge 
correspondence~\cite{Wen_PRL90,Hatsugai_93,Hatsugai_PRB93}, which is a 
universal feature 
of topological phases~\cite{Kane_PRL05,Haldane_PRL08,Hasan_PMP10,Qi_PMP11,Kariyado_SR15,Delplace_Science17,Sone_PRL19,Hatsugai_PRB16,Kuno_PRR20,Mizoguchi_PRB21,Yoshida_arXiv20}. 
The edges of the QH systems demonstrate the nontrivial transport 
properties enriched by the bulk topology, which has attracted a great interest
for over decades~\cite{Laughlin_PRL81,Halperin_82,MacDonald_PRL90,Wen_Luttinger_PRB90,Wen_PRL90,Wen_PRB91,MacDonald_PRL91,Wen_MPB92,Wen_PRB94,Wen_AP95,Meir_PRL94,Kane_PRL94,Kane_PRB95,Rezayi_PRL02,Joglekar_PRB03,Rezayi_PRB03,Chang_RMP03,Rezayi_PRB08,Rezayi_PRB09,Jianhui_PRL13,Neupert_NPJ18,Simon_PRB18,Wei_PRB20,Shibata_PRB21,Khanna_PRB21}.
The main goal in this work is to reveal how the quantum Hall edge states are
evolved during the process of the flux-attachment in the adiabatic heuristic
principle.

In this Letter, we analyze the fractional pumping phenomena associated 
with the Laughlin's argument of the anyonic FQH effect. We show that the 
general feature of the energy spectrum with edges shows little change during
the process of the flux-attachment while the topological degeneracy is wildly 
changed. Furthermore, the total jump of the center-of-mass works well as an 
invariant of this process, which is consistent with the bulk-edge 
correspondence of the FQH effect of anyons. This implies that the total jump of
the center-of-mass characterizes the fractional charge pumping of the
adiabatic heuristic principle. Also, this supports the stability of 
the QH edge states in the adiabatic heuristic principle.

\begin{figure}[b!]
  \begin{center}  
   \includegraphics[width=\columnwidth]{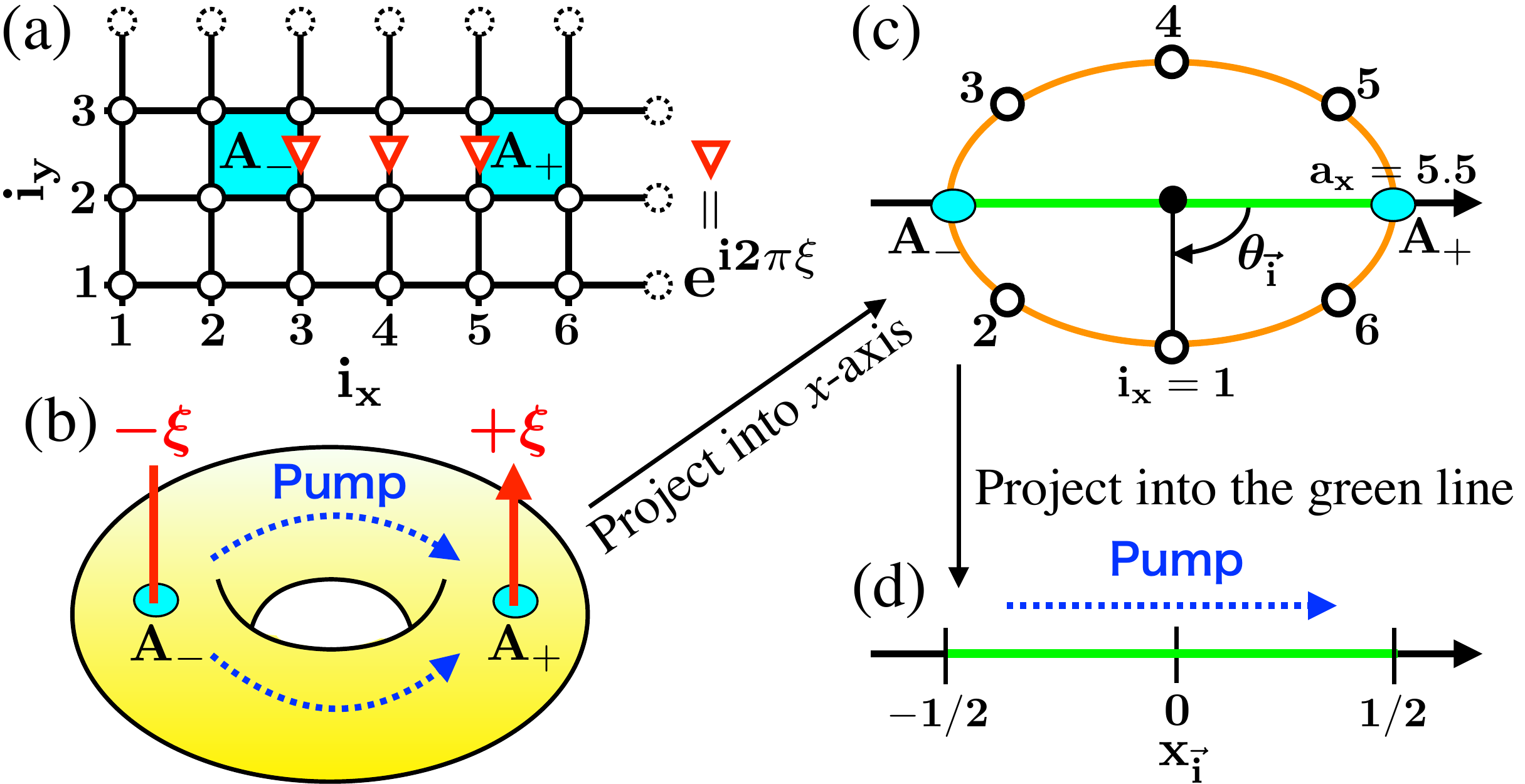}
  \end{center}
 \caption{
 (a) Sketch of $6\times3$ square lattice. The gauge $\xi_{ij}$
 (red arrows) describes the two local fluxes $\pm\xi$ at $A_{\pm}$. 
 (b) Charge pump from $A_-$ to $A_+$. 
 (c) One-dimensional projection into $x$-axis. The projected 
 sites for (a) are shown. The angle $\theta_{\vec{i}}$ is measured from $a_x$
 that is the $x$ coordinate of $A_+$.
 (d) One-dimensional charge pump on $-1/2\leq x_{\vec{i}}\leq1/2$.
 }
 \label{fig:solenoid}
\end{figure}

\paragraph{Charge pumping}---
Let us consider the QH system on a square 
lattice with $N_x\times N_y$ sites, where $N_x/N_y=2$ and the periodic boundary
condition is imposed. As shown in Fig.~\ref{fig:solenoid}(a), local fluxes 
$\pm\xi$ are set at two plaquettes $A_{\pm}$ with the same $y$ coordinate.
Their distance is $N_x/2$. Particles are pumped from $A_{-}$ to $A_{+}$ 
as $\xi$ varies from $0$ to $1$ [see Fig.~\ref{fig:solenoid}(b)], which we call
the (fractional) charge 
pump~\cite{Thouless_PRB83,Dai_PRL13,Nakajima_Nature16,Lohse_Nature16,Guo_PRB12,Chen_PRL13,Zeng_PRL15,Chen_PRB16,Sheng_PRB16,Nakagawa_PRB17,Taddia_PRL17,Nakagawa_PRB18}
throughout this Letter.

This charge pump can be mapped into the one-dimensional pump with {\it edges} 
[Fig.~\ref{fig:solenoid}(d)]. As shown in Fig.~\ref{fig:solenoid}(c), we first 
project the system into the $x$-axis. Then, projecting it into the
green line shown in Figs~\ref{fig:solenoid}(c), we finally define a new 
coordinate for site 
$\vec{i}=(i_x,i_y)$ as $x_{\vec{i}}=(1/2)\cos\theta_{\vec{i}}$ with 
$\theta_{\vec{i}}=2\pi(i_x-a_x)/N_x$, where $a_x$ is the $x$ coordinate of 
$A_{+}$, see Fig.~\ref{fig:solenoid}(d). In this projection, the two pin-holes 
$A_\pm$ are mapped into the edges $x_{\vec{i}}=\pm1/2$.

The charge can be transformed from $x_{\vec{i}}=-1/2$ to 
$x_{\vec{i}}=1/2$ as $\xi$ increases. The pumped charge is given by the 
integration of 
$\pa_\xi P(\xi)$, where $P$ is the center-of-mass,
\begin{align}
 P(\xi)=\text{Tr}\,[\rho(\xi)\sum_{\vec{i}}x_{\vec{i}}n_{\vec{i}}],
 \label{eq:cm}
\end{align}
$\rho$ is the zero temperature density matrix in the grand canonical 
ensemble and $n_{\vec{i}}$ is the number operator at the site $\vec{i}$. 
(In Sec.~\ref{Sec:current} of Supplemental Material~\cite{sup}, we derive the 
pumped charge by using the current operator.)
As $\xi$ varies, $P(\xi)$ jumps several times due to the sudden
change of the particle number~\cite{Hatsugai_PRB16,Kuno_PRR20,Kuno_arXiv21}. 
Accordingly, the pumped
charge between the period $\xi\in[0,1]$ is given by
$Q=\left(\int_0^{\xi_1^-}+\int_{\xi_1^+}^{\xi_2^-}+\cdots
+\int_{\xi_N^+}^{1}\right)d\xi\,\pa_\xi P(\xi)$, where
$\xi_1,\cdots,\xi_N$ are the jumping points in the period and 
$\xi_\alpha^\pm=\xi_\alpha\pm0$. Using the periodicity $P(1)=P(0)$ and 
$\Delta P(\xi_\alpha)\equiv P(\xi_\alpha^+)-P(\xi_\alpha^-)$, we 
get~\cite{Hatsugai_PRB16}
\begin{align}
 Q=-\sum_{\alpha=1}^N\Delta P(\xi_\alpha)\equiv-\Delta P_\text{tot}.
 \label{eq:deltaQ}
\end{align}
As shown below, the total jump $\Delta P_\text{tot}$, i.e., the sudden changes 
of the particle number, comes from the (dis)appearance of edge states. 
Equation~\eqref{eq:deltaQ} implies that the pumped charge is 
given only by the information of edges. 

\paragraph{Bulk-edge correspondence}---
In this Letter, we numerically show the following bulk-edge 
correspondence for the FQH states of anyons: 
\begin{align}
 C=-N_D\times\Delta P_\text{tot},
 \label{eq:BEC}
\end{align}
where $C$ is the many-body Chern number~\cite{Niu_PRB85} of the $N_D$-fold 
degenerate ground state multiplet at $\xi=0$.
This is consistent with the Laughlin's argument applied to the FQH 
systems~\cite{Laughlin_PRL81,Halperin_82,Thouless_PRL82,Thouless_PRB83,Niu_PRB85,Hatsugai_93,Thouless_PRB89,Hatsugai_PRB16,Sheng_PRB16,Pollman_PRB15,Andrews_PRB21} that 
implies $Q=C/N_D$. In the following, we clarify how the fractional charge 
pumping is deformed to the standard pumping phenomena by the flux-attachment
transformation. As mentioned below, the relation in Eq.~\eqref{eq:BEC} results 
in the stability of the QH edge states in the adiabatic heuristic principle.

\begin{figure}[t!]
  \begin{center}  
   \includegraphics[width=\columnwidth]{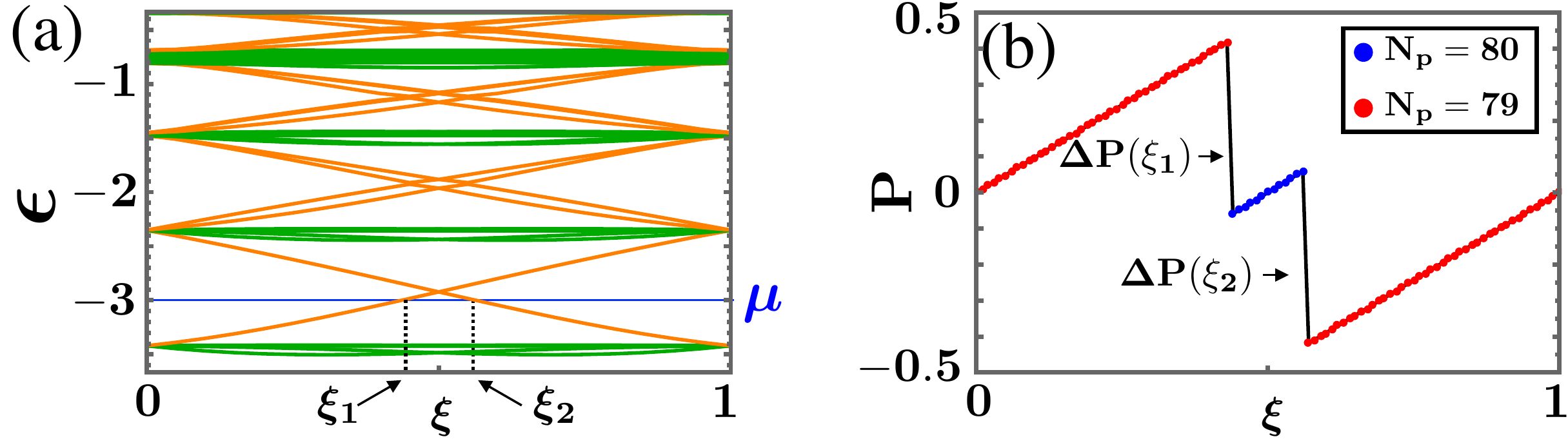}
  \end{center}
 \caption{
 (a) Single-particle energy $\epsilon$ on $40\times20$ lattices with 
 $\phi=1/10$. The green (orange) plots represent the bulk (edge) states. The 
 blue line is $\mu=-3$.
 (b) Center-of-mass $P$. The two jumps are 
 $\Delta P(\xi_1)\approx\Delta P(\xi_2)\approx-0.48$.
 }
 \label{fig:nonint}
\end{figure}

\paragraph{Fermion pumping}---
As a first step, we confirm Eq.~\eqref{eq:BEC} for the IQH system of 
non-interacting fermions. The Hamiltonian is 
$H=-t\sum_{\langle ij\rangle}e^{i\phi_{ij}}e^{i\xi_{ij}}c_i^\dagger c_j$, where
$c_i^\dagger$ is the creation operator for a fermion on site $i$ and $t=1$. The
phase factors $e^{i\phi_{ij}}$ and $e^{i\xi_{ij}}$ describe the uniform 
magnetic field~\cite{Hofstadter_PRB76,Hatsugai_PRL99} and the local fluxes 
at $A_{\pm}$ [see Fig.~\ref{fig:solenoid}(a)], respectively. 
We plot in Fig.~\ref{fig:nonint}(a) the single-particle energy $\epsilon$
with $N_x\times N_y=40\times20$ and $\phi\equiv N_\phi/(N_xN_y)=1/10$, where 
$N_\phi$ is the total uniform fluxes. Each set of $N_\phi(=80)$ states forms 
the Landau level (LL) at $\xi=0$. As $\xi$ increases, some edge states go over 
to the mid-gap region. In Fig.~\ref{fig:nonint}(b), we compute $P(\xi)$
with $\rho=\ket{G}\bra{G}$ in Eq.~\eqref{eq:cm}, where $\ket{G}$ is the 
ground state completely occupying the 1st LL under the chemical potential 
(Fermi energy) $\mu=-3$. The sudden change of the particle number $N_p$ causes
the jumps of $P(\xi)$ at $\xi_1$ and $\xi_2$. Both values of 
$\Delta P(\xi_\alpha)$'s are 
approximately $-1/2$, which is consistent 
with Fig.~\ref{fig:nonint}(a) where one edge state at $x_{\vec{i}}=1/2$ goes 
over across $\mu$ and then another at $x_{\vec{i}}=-1/2$ goes back. Although 
a finite size effect gives
$\Delta P_\text{tot}=\Delta P(\xi_1)+\Delta P(\xi_2)\approx-0.96$, we confirm 
$\Delta P_\text{tot}=-1$ in the thermodynamic limit 
(see Sec.~\ref{Sec:finite} of Supplemental Material~\cite{sup}).
This is consistent with Eq.~\eqref{eq:BEC} with $N_D=1$ and $C=1$. The cases 
for $C=2$ and $3$ have been also confirmed.

\paragraph{Normalized jumps}---
As mentioned above, the jump $\Delta P(\xi_\alpha)$ are not quantized to
$\pm1/2$ due to the finite size effect. Let us then properly normalize each
jumps: when $P$ jumps positively or negatively at $\xi_\alpha$, we assign it as
$\Delta P(\xi_\alpha)\mapsto1/2$ or $-1/2$.
Hereafter ``$\mapsto$'' denotes this normalization; e.g., we have
$\Delta P_\text{tot}=P(\xi_1)+P(\xi_2)\mapsto-1/2-1/2=-1$ in 
Fig.~\ref{fig:nonint}(b). This 
gives the bulk-edge correspondence in Eq.~\eqref{eq:BEC} even for finite 
systems.

\paragraph{Fractional anyon pumping}---
\begin{figure*}[t!]
   \includegraphics[width=2\columnwidth]{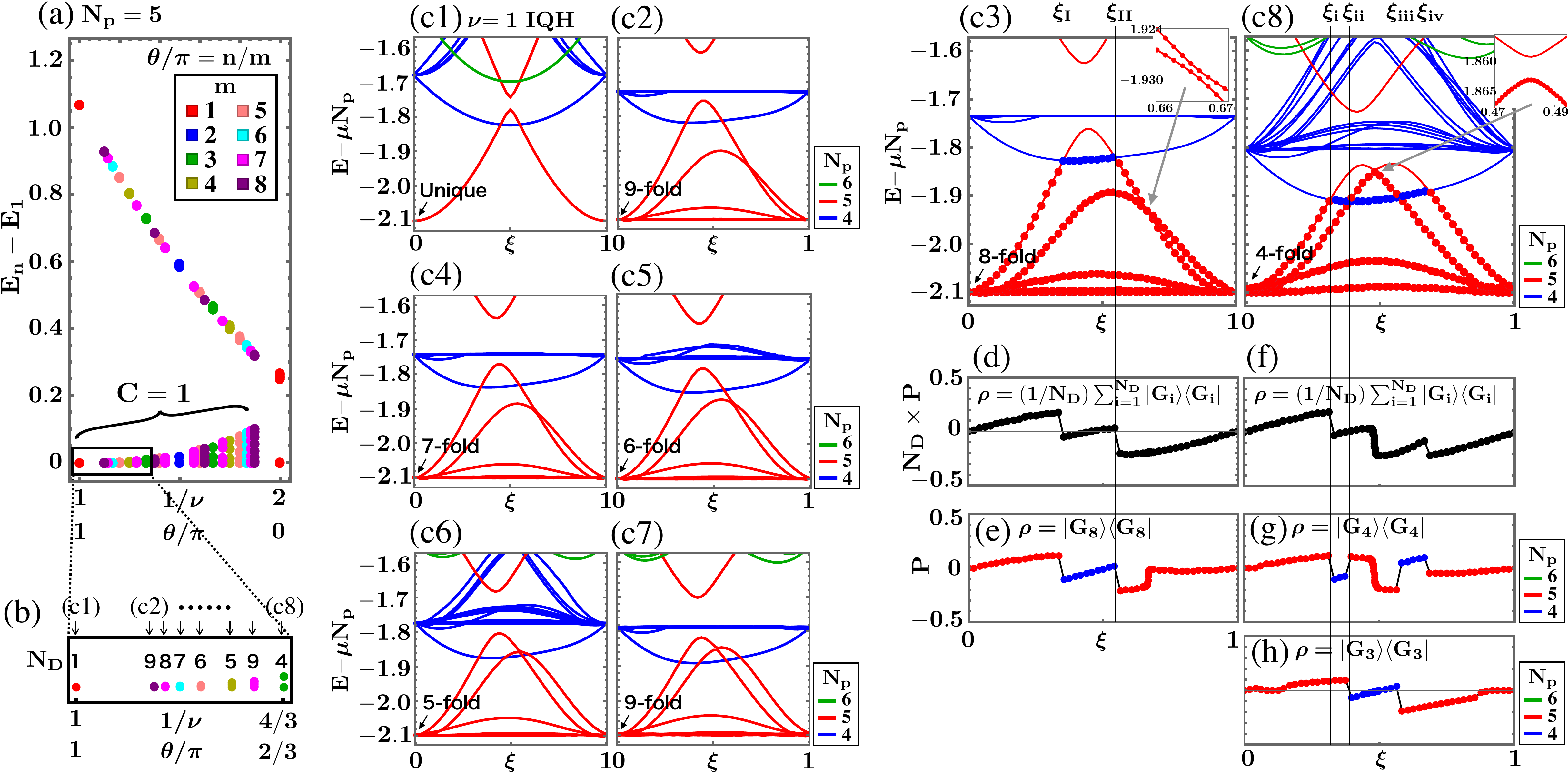}
 \caption{
 (a) Energy gaps as functions of $1/\nu$ at $\xi=0$ with $N_p=5$. The 
 statistical angle $\theta$ is  determined by $\nu=p/[p(1-\theta/\pi)+1]$ with 
 $p=1$. The color expresses the denominator of $\theta/\pi$. 
 (b) Ground state degeneracy $N_D$. 
 (c1)-(c8) Energy spectra as functions of local fluxes $\xi$. Each value of 
 ($\nu,\theta)$ is represented above the panel (b). We set $\mu=-3$ in 
 (c1) and choose $\mu$ in (c2)-(c8) so that the lowest energy at $\xi=0$ is
 same as that in (c1). In (c3) and (c8), the $N_D$ lowest energy states are
 marked by dots.
 In (a)-(c), we plot the lowest $N_\text{cut}$ energies with $N_\text{cut}=2$ 
 for $N_p=6$ and $N_\text{cut}=20$ for $N_p=5,4$.
 (d)-(h) Center of mass $P$ as for the panels (c3) and (c8).
 $\xi_\text{I}$,$\xi_\text{II}$ and $\xi_\text{i}$-$\xi_\text{iv}$ represent 
 the gap closing points. 
 }
 \label{fig:ah}
\end{figure*}
Let us consider the fractional pumping of anyons. To this end, we
take the Hamiltonian as 
$H=-t\sum_{\langle ij\rangle}e^{i\phi_{ij}}e^{i\xi_{ij}}e^{i\theta_{ij}}c_i^\dagger c_j$, where the phase factor 
$\theta_{ij}$~\cite{Wen_anyon_PRB90,Hatsugai_PRB91,fermi}
depends on the configuration of all particles $\{\bm{r}_k\}_{1\leq k\leq N_p}$,
which describes the fractional statistics $e^{i\theta}$. 
Note that although $c_i^\dagger$ is the creation operator for a fermion, $H$ is
the Hamiltonian of anyons and 
includes intrinsically the many-body interactions. Due to 
constraints of the braid group, $\dim H$ depends on $\theta$ even for the same 
$N_p$~\cite{Wen_anyon_PRB90,Einarsson_PRL90}: the Hilbert space for 
$\theta/\pi=n/m$ ($n,m$: coprime) is spanned by the basis 
$\ket{\{\bm{r}_k\};w}$, where 
$w=1,\cdots,m$ is an additional 
internal degree of freedom. When a particle hops across the boundary in 
the $x$ direction, the label is shifted from $w$ to $w-1$. 
As for the boundary in the $y$ direction, the phase factor $e^{iw\theta}$ 
is given.
Thanks to this, global requirements of anyons 
hold~\cite{GR,Wen_anyon_PRB90,Einarsson_PRL90,Hatsugai_PRB91}. Also, we 
introduce $\xi_{ij}$ only for the basis with $w=1$~\cite{defC}.

In the following, we focus on a family of the $\nu=1$ IQH states connected by 
trading the magnetic fluxes for statistical 
ones~\cite{Jain_CFT_PRL89,Greiter_NPB90,Greiter_NPB92,Kudo_PRB20}:
$\nu=p/[p(1-\theta/\pi)+1]$ with $p=1$. Fixing $N_x\times N_y=10\times5$, 
$N_p=5$ and $\xi=0$, we plot the energy gaps as functions of $1/\nu$ in 
Fig.~\ref{fig:ah}(a). 
Due to the lattice, the topological degeneracy is lifted. We here define the 
low-energy states with $E_n-E_1<0.2$ as the 
ground state multiplet. 
The ground state at $\nu=s/t$ ($s,t$: coprime) in Fig.~\ref{fig:ah}(a) 
gives the degeneracy $N_D=t$ numerically [see Fig.~\ref{fig:ah}(b)] and the 
Chern numbers of the multiplets are always $C=1$~\cite{Kudo_PRB20}.
Namely, the many-body Chern number is 
used as an adiabatic invariant. The gap closing at $\nu\approx1/2$ is 
expected due to finite-size effects~\cite{Kudo_PRB20}. 

Let us investigate the pumping phenomena. As for each parameter 
$(1/\nu,\theta/\pi)$ shown in Fig.~\ref{fig:ah}(b), we plot in 
Figs.~\ref{fig:ah}(c1)-(c8) the eigenvalues of 
the Hamiltonian including the chemical potential, $H-\mu N_p$, as functions of 
$\xi$ with $4\leq N_p\leq6$~\cite{Np}. Figure~\ref{fig:ah}(c1) is in the 
same setting of Fig.~\ref{fig:nonint} but for smaller system sizes. In 
Fig~\ref{fig:ah}(c1), the particle number $N_p$ of the unique ground state is 
changed as $\xi$ increases due to the (dis)appearance of the edge state
as mentioned previously. The gap between the two red lines at $\xi\approx0.5$ 
is a finite-size effect. As shown in Figs.~\ref{fig:ah}(c1)-(c8), even though 
the topological degeneracy is wildly changed as $1/\nu$ and $\theta/\pi$ vary, 
the general feature of the spectra remains unchanged. The degenerate 
ground states at $\xi=0$ are lifted as $\xi$ increases and then one or two 
states float up in energy to cross with another state having one particle less.

Now we focus on the anyonic system in Fig.~\ref{fig:ah}(c3) 
and show its the bulk-edge correspondence. Here $\theta/\pi=6/7$, 
$\nu=7/8$ and $N_D=8$ at 
$\xi=0$. To define the center-of-mass of
the ground state multiplet suitably, we define the density 
matrix as $\rho(\xi)=(1/N_D)\sum_{k=1}^{N_D}\ket{G_k(\xi)}\bra{G_k(\xi)}$, 
where 
$\ket{G_k(\xi)}$ is the $k$-th lowest energy state. Using it with 
Eq.~\eqref{eq:cm}, we plot $N_D\times P(\xi)$ in Fig.~\ref{fig:ah}(d). There 
are two jumps at $\xi_\text{I}$ and $\xi_\text{II}$, where the $N_D$th and 
$N_D+1$th lowest energy states cross each other in the spectrum. Because of 
$N_D\rho=\sum_{k=1}^{N_D}\ket{G_k}\bra{G_k}$, the obtained jumps are 
solely given by $P$ with $\rho=\ket{G_{N_D}}\bra{G_{N_D}}$ shown in 
Fig.~\ref{fig:ah}(e). This figure gives
$\Delta P_\text{tot}=\Delta P(\xi_I)+P(\xi_{II})\mapsto-1$, which implies 
$N_D\times\Delta P_\text{tot}\mapsto-1$ in Fig.~\ref{fig:ah}(d). 
This is consistent with Eq.~\eqref{eq:BEC} with $C=1$. 
Because of $Q=-\Delta P_\text{tot}$, we ahve the fractional pumped charge 
$Q=1/8$.
In this argument, we assume the 
absence of 
the gap closing between states with the same $N_p$ apart from $\xi=0$ since 
there are no symmetry except for the charge
$U(1)$.
The gap at $\xi\approx0.7$ between $\ket{G_{N_D}}$ and 
$\ket{G_{N_D-1}}$ is very small but is finite~\cite{TwoEdge} as shown in the 
inset.

\begin{figure}[t!]
  \begin{center}  
   \includegraphics[width=\columnwidth]{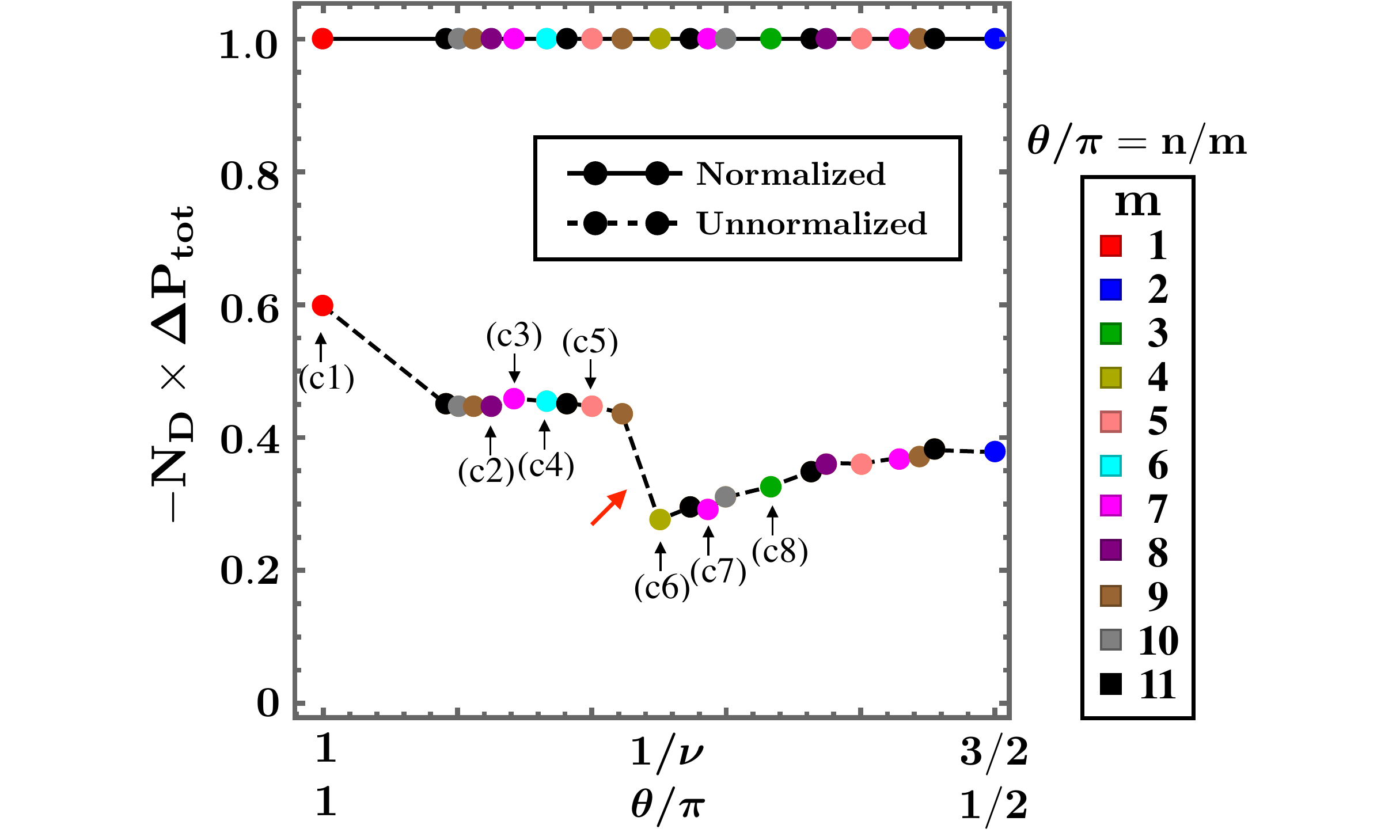}
  \end{center}
 \caption{
 The total jump of the center-of-mass $N_D\times\Delta P_\text{tot}$. The 
 solid and dotted lines mean the normalized and the unnormalized data, 
 respectively. The systems in Figs.~\ref{fig:ah}(c1)-(c8) are marked. The 
 sudden change in the dotted line, represented by a red arrow, is due to the 
 gap collapse of the $N_D-1$th lowest energy state, compare 
 Figs.~\ref{fig:ah}(c5) and \ref{fig:ah}(c6). 
 The jumps are calculated by discretizing the period $\xi\in[0,1]$ into
 $N_\xi$ meshes with $N_\xi=48$ (Only at (c6) point, we set $N_\xi=240$).
 }
 \label{fig:invariant}
\end{figure}

Let us here mention the finite size effect in Fig.~\ref{fig:ah}(d). The value 
of $N_D\times\Delta P_\text{tot}$ before normalizing is about $-0.46$, which is
far away from $-1$. Although this value in the IQH system in 
Fig.~\ref{fig:ah}(c1) is about the same magnitude (about $-0.60$, see the data
point at $\nu=1$ of Fig.~\ref{fig:invariant}), it approaches $-1$ as the system
size increases as
confirmed in Sec.~\ref{Sec:finite} of Supplemental Material~\cite{sup}. Since
the bulk gaps of the two systems are comparable and their system sizes are 
same, the deviation from $-1$ in Fig.~\ref{fig:ah}(d) is also expected to be
the finite size effect. 

Let us next focus on the system in Fig.~\ref{fig:ah}(c8), where 
$\theta/\pi=2/3$, $\nu=3/4$ and $N_D=4$ at $\xi=0$. Unlike the previous case, 
there are four gap-closing points, $\xi_\text{i}\cdots\xi_\text{iv}$, as for
the $N_D$ lowest energy states. However, 
$P(\xi)$ with $\rho=(1/N_D)\sum_{k=1}^{N_D}\ket{G_k}\bra{G_k}$ 
in Fig.~\ref{fig:ah}(f) jumps only at $\xi_\text{i}$ and $\xi_\text{iv}$
because the jumps at $\xi_\text{ii},\xi_\text{iii}$ cancel each other, see 
Figs.~\ref{fig:ah}(g) and 
\ref{fig:ah}(h). Consequently, the total jump is given by 
$N_D\times\Delta P_\text{tot}\mapsto-1$, which is consistent with 
Eq.~\eqref{eq:BEC} with $C=1$. 
This implies the fractional pumped charge $Q=1/4$.
The gap at $\xi\approx0.5$ is very small but is finite as mentioned 
before, see the inset in Fig.~\ref{fig:ah}(c8).

The results shown in Figs.~\ref{fig:ah}(c3) and \ref{fig:ah}(c8) 
suggest that $N_D\times\Delta P_\text{tot}$ is 
the invariant of the bulk gap
in the process of the flux-attachment. To demonstrate it, we plot the total 
jumps
as functions of $1/\nu$ in Fig.~\ref{fig:invariant}, where both data 
before/after normalizing each jumps are shown. The normalized data justify that
$N_D\times\Delta P_\text{tot}$ works well as 
the invariant. This nature is also 
indicated by the unnormalized data 
in Fig.~\ref{fig:invariant}: the plots are
{\it smooth} as $1/\nu$ and $\theta/\pi$ vary even though (i) the degeneracy 
$N_D$ is {\it wildly} changed and 
(ii) the dimension of the Hamiltonian is {\it discretely} changed depending on 
the denominator of $\theta/\pi$: e.g., 
with $N_p=5$, $\dim H=\binom{N_xN_y}{N_p}=\binom{50}{5}=2118760$ for 
$\theta/\pi=n$ while $\dim H=11\binom{50}{5}=23306360$ for $\theta/\pi=n/11$
(this is due to the additional 
internal degree $w$ of the basis 
$\ket{\{\bm{r}_k\};w}$ as mentioned above). We stress that this nontrivial 
smoothness in Fig.~\ref{fig:invariant} implies the stability of the QH edge 
states in the adiabatic heuristic principle.

\paragraph{Conclusion}---
In this Letter, we demonstrate the bulk-edge correspondence of the FQH states 
of anyons. 
The results indicate that the total jump of the 
center-of-mass, which corresponds to the many-body Chern number, is an 
invariant with respect to the flux-attachment. This implies the 
stability of edge states in the adiabatic heuristic principle.
Recently, direct observation of the center of mass in pumping phenomena 
has been conducted in cold atoms~\cite{Nakajima_Nature16,Lohse_Nature16}. 
The behavior of the center-of-mass that we focus on would be observed in cold 
atoms although the experimental realization of the two-dimensional anionic 
system is still a challenging problem.

\acknowledgements
We thank the Supercomputer Center, the Institute for Solid State Physics, the 
University of Tokyo for the use of the facilities. The work is supported in 
part by JSPS KAKENHI Grant Numbers JP17H06138, JP19J12317 (K.K.), and
JP21K13849 (Y.K.).

\bibliographystyle{apsrev4-1}
\bibliography{citation}

\begin{thebibliography}{95}
\bibitem [{fer()}]{fermi}%
  \BibitemOpen
  \href@noop {} {}\bibinfo {note} {The Hamiltonian is also expressed by using
  the creation-annihilation operators of hard-core bosons with the statistical
  flux. In the numerical calculations, we diagonalize it because of its
  technical simplicity.}\BibitemShut {Stop}%
\bibitem [{GR()}]{GR}%
  \BibitemOpen
  \href@noop {} {}\bibinfo {note} {A global requirement of anyons on a
  torus~\cite{Einarsson_PRL90,Wen_anyon_PRB90} is
  $\rho_i^{-1}\tau_j\rho_i\tau_j^{-1}=A_{ij}$, where $\tau_i$ and $\rho_i$ are
  global move operators of particle $i$ along noncontractible loops on the
  torus in $x$ and $y$ directions, and $A_{ij}(=e^{i2\theta})$ is a local move
  operator of particle $i$ around particle $j$.}\BibitemShut {Stop}%
\bibitem [{def()}]{defC}%
  \BibitemOpen
  \href@noop {} {}\bibinfo {note} {The many-body Chern number that is invariant
  for the adiabatic heuristic principle is given by the following twisted
  boundary conditions~\cite{Kudo_PRB20}: when an anyon hops across the boundary
  in $x$ ($y$) direction, the phase factor $e^{i\theta_x\delta_{w,1}}$
  ($e^{i\theta_y}$) is given to the basis $\ket{\{\bm{r}_k\};w}$. According to
  this rule in $x$ direction, we define the local fluxes $\pm\xi$ only for
  $w=1$.}\BibitemShut {Stop}%
\bibitem [{Np()}]{Np}%
  \BibitemOpen
  \href@noop {} {}\bibinfo {note} {For fermions, we change $N_p$ keeping
  $N_\phi$ constant. Such a change for anyons is prohibited by constraints of
  the braid group~\cite{Einarsson_PRL90}. Thus we change their $N_p$ in such a
  way that the total flux $N_\phi+N_p\times\theta/\pi$ is same as that in the
  case of fermions.}\BibitemShut {Stop}%
\bibitem [{Two()}]{TwoEdge}%
  \BibitemOpen
  \href@noop {} {}\bibinfo {note} {Because of the mid-gap region,
  $\ket{G_{N_D}}$ and $\ket{G_{N_D-1}}$ at $\xi\approx0.7$ have edge modes. It
  is suggested that they are localized at the opposite edges each other. Either
  one is expected due to finite size effects.}\BibitemShut {Stop}%
\end{thebibliography}
\clearpage

\renewcommand{\thesection}{S\arabic{section}}
\renewcommand{\theequation}{S\arabic{equation}}
\renewcommand{\thefigure}{S\arabic{figure}}
\renewcommand{\thetable}{S\arabic{table}}
\setcounter{equation}{0}
\setcounter{figure}{0}
\setcounter{table}{0}
\setcounter{page}{1}
\makeatletter
\c@secnumdepth = 2
\onecolumngrid
\begin{center}
 \Large{Supplemental Material}
\end{center}
\twocolumngrid

\section{Current operator and the pumped charge}
\label{Sec:current}
In this appendix, we derive the pumped charge $Q$ by using the current 
operator. We consider the Hamiltonian described in the paragraph 
{\it Fractional anyon pumping} in the main text:
\begin{align}
 H(\xi)
 =-t\sum_{\langle \vec{i},\vec{j}\rangle}e^{i\phi_{\vec{i},\vec{j}}}
 e^{i\xi_{\vec{i},\vec{j}}}e^{i\theta_{\vec{i},\vec{j}}}
 c^\dagger_{\vec{i}}c_{\vec{j}},
 \label{eq:ham}
\end{align}
where $c_{\vec{i}}^\dagger$ is the creation operator for a fermion on site 
$\vec{i}$, and the phase factors $e^{i\phi_{\vec{i},\vec{j}}}$, 
$e^{i\xi_{\vec{i},\vec{j}}}$ and $e^{i\theta_{\vec{i},\vec{j}}}$ describe the 
uniform fluxes, the local fluxes $\pm\xi$ and the statistical phase
$\theta$, respectively. Using a unitary operator,
\begin{align*}
 U(\alpha)=\prod_{\vec{i}}e^{-i\alpha x_{\vec{i}}n_{\vec{i}}},
\end{align*}
let us modify the Hamiltonian in Eq.~\eqref{eq:ham}:
\begin{align}
 H(\alpha,\xi)
 &\equiv U(\alpha)H(\xi)U^\dagger (\alpha)\nonumber\\
 &=-t\sum_{\langle \vec{i},\vec{j}\rangle}
 e^{-i\alpha(x_{\vec{i}}-x_{\vec{j}})}
 e^{i\phi_{\vec{i},\vec{j}}}
 e^{i\xi_{\vec{i},\vec{j}}}e^{i\theta_{\vec{i},\vec{j}}}
 c^\dagger_{\vec{i}}c_{\vec{j}},
 \label{eq:gauge}
\end{align}
where 
$U(\alpha)c_{\vec{i}}U^\dagger(\alpha)=
e^{i\alpha x_{\vec{i}}}c_{\vec{i}}$ is used.
We then define the current operator in the $x$ direction as
\begin{align*}
 \mathcal{I}_x
 &=i\frac{t}{\hbar}
 \sum_{\langle \vec{i},\vec{j}\rangle}(x_{\vec{i}}-x_{\vec{j}})
 e^{-i\alpha(x_{\vec{i}}-x_{\vec{j}})}
 e^{i\phi_{\vec{i},\vec{j}}}
 e^{i\xi_{\vec{i},\vec{j}}}
 e^{i\theta_{\vec{i},\vec{j}}}
 c^\dagger_{\vec{i}}c_{\vec{j}}\nonumber\\
 &=\frac{1}{\hbar}\pa_\alpha H(\alpha,\xi).
\end{align*}
We now assume the following density matrix:
\begin{align*}
 \rho(\alpha,\xi)=\frac{1}{N_D}\Phi(\alpha,\xi)\Phi^\dagger(\alpha,\xi),
\end{align*}
where 
$\Phi(\alpha,\xi)=(\ket{G_1(\alpha,\xi)},\cdots,\ket{G_{N_D}(\alpha,\xi)})$ is 
the ground state multiplet of the Hamiltonian $H(\alpha,\xi)$. The measured
current $I_x$ computed from $\mathcal{I}_x$ is reduced to the Berry 
curvature~\cite{Thouless_PRB83,Hatsugai_PRB16}:
\begin{align*}
 &I_x(\alpha,\xi)=-\frac{i}{N_D}B(\alpha,\xi),\\
 &B(\alpha,\xi)=\pa_\alpha A_\xi(\alpha,\xi)-\pa_\xi A_\alpha(\alpha,\xi),\\
 &A_{s}(\alpha,\xi)=
 \text{tr}\left[\Phi^\dagger(\alpha,\xi)\pa_s\Phi(\alpha,\xi)\right],
 \ s=\alpha,\xi,
\end{align*}
where ``tr'' indicates the trace of a $N_D$-dimensional matrix. The gauge 
transformation in Eq.~\eqref{eq:gauge} implies 
$\Phi(\alpha,\xi)=U(\alpha)\Phi(\xi)$, i.e.,
\begin{align*}
 &A_\alpha(\alpha,\xi)
 =-\text{tr}\,\left[
 \Phi(\xi)^\dagger\left(\sum_{\vec{i}}ix_{\vec{i}}n_{\vec{i}}\right)
 \Phi(\xi)
 \right]
 =iN_DP(\xi),\\
 &A_\xi(\alpha,\xi)
 =\text{tr}\,\left[
 \Phi(\xi)^\dagger\pa_\xi \Phi(\xi)
 \right]
 \equiv A_\xi(\xi),
\end{align*}
where $P(\xi)$ is the center-of-mass defined in Eq.~\eqref{eq:cm}. By fixing
the gauge $\Phi(\xi)$ properly~\cite{Thouless_PRB83,Hatsugai_JPSJ04,Hatsugai_PRB16}, one can prepare 
$A_\xi(\xi)$ as a well-defined function. Because of 
$\pa_\alpha A_\xi(\xi)=0$, we have $B(\alpha,\xi)=iN_D\pa_\xi P(\xi)$, namely,
\begin{align}
 I_x(\alpha,\xi)=\pa_\xi P(\xi).
\end{align}
The pumped charge $Q$ is given by the integration of the current 
$I_x(\alpha,\xi)$ over $\xi$. This clearly justifies the formulation using 
$P(\xi)$ in the main text.

\section{Finite size effect}
\label{Sec:finite}
\begin{figure}[b!]
  \begin{center}  
   \includegraphics[width=\columnwidth]{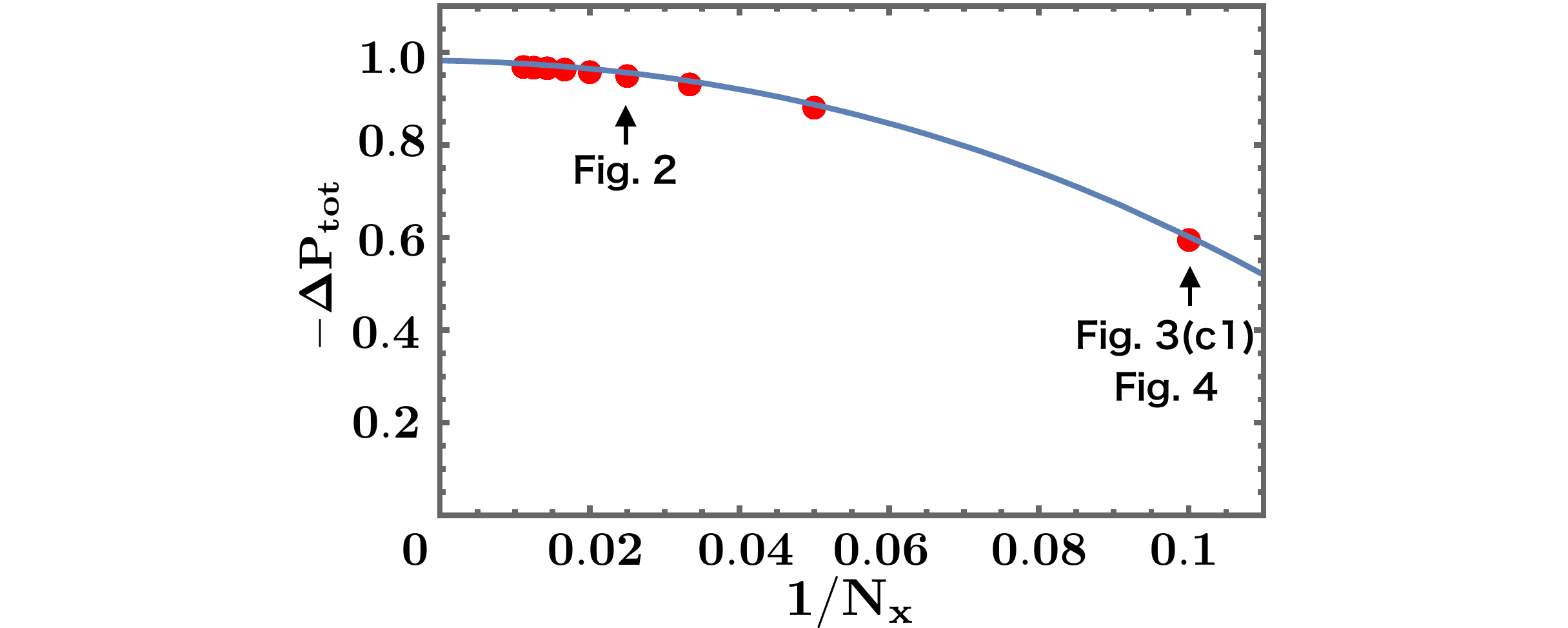}
  \end{center}
 \caption{
 Finite-size scaling analysis of the total jump $\Delta P_\text{tot}$ at 
 $\nu=1$. The parameters are fixed as $N_x/N_y=2$, $\mu=-3$, and $\phi=1/10$. 
 Toward $N_x\rightarrow\infty$, the data extrapolate to 
 $\Delta P_\text{tot}=-0.98$.
 }
 \label{fig:finite}
\end{figure}

Let us discuss a finite size effect of numerical simulation. In 
Fig.~\ref{fig:finite}, we plot the total jump $\Delta P_\text{tot}$ at $\nu=1$,
where we fix the aspect ratio, the chemical potential, and flux per plaquette 
as $N_x/N_y=2$, $\mu=-3$, and $\phi=1/10$, respectively. As indicated in the 
figure, the systems
in Figs.~\ref{fig:nonint}, \ref{fig:ah}(c1) and \ref{fig:invariant} are 
included. As the system
size $N_x$ increases, $\Delta P_\text{tot}$ approaches -1. This implies that
the edge modes are somewhat spread over in a finite size system, which results
in the deviation from $-1$ at $\nu=1$ in Figs.~\ref{fig:nonint} and 
\ref{fig:invariant}.

\end{document}